\documentstyle[preprint,aps]{revtex}
\begin{document}

\title{S-, P- and D-wave resonances in positronium-sodium and 
positronium-potassium scattering} 

\author{Sadhan K Adhikari$^\dagger$ and Puspajit Mandal$^{\dagger \$}$}
\address{$^\dagger$Instituto de F\'{\i}sica Te\'orica, 
Universidade Estadual Paulista, 
01.405-900 S\~ao Paulo, S\~ao Paulo, Brazil\\
$^{\$}$Department of Mathematics, Visva Bharati, Santiniketan 731 235,
India}

\date{\today}
\maketitle

\begin{abstract}

Scattering of positronium (Ps)  by sodium and potassium atoms  has been
investigated employing a three-Ps-state coupled-channel model with
Ps(1s,2s,2p) states using a  time-reversal-symmetric regularized
electron-exchange model potential fitted to reproduce accurate
theoretical  results for PsNa and PsK binding energies.  
We find a narrow S-wave singlet resonance at 4.58 eV of width 0.002 eV in
the
Ps-Na system and at 4.77 eV of width 0.003 eV in the Ps-K system.
Singlet P-wave
resonances in both systems are found at 5.07 eV of width 0.3 eV. Singlet 
D-wave structures are  found  at 5.3 eV  in both systems.
We also report results for elastic and Ps-excitation cross sections for Ps
scattering by Na and K.

{\bf PACS Number(s):  34.10.+x, 36.10.Dr}

\end{abstract}


\newpage

Recent successful high precision measurements of positronium
(Ps) scattering by H$_2$, N$_2$, He, Ne, Ar, C$_4$H$_{10}$, and
C$_5$H$_{12}$ \cite{1,2} have enhanced theoretical activities
\cite{3a,18,5,5x} in this subject. 
We suggested \cite{6} a regularized, symmetric,  nonlocal
electron-exchange
model potential and used it  
in the
successful study of Ps scattering by H \cite{7}, He \cite{6,6a,6x,15},
Ne
\cite{15}, Ar \cite{15}, H$_2$ \cite{16} and Li \cite{15a}. Our results
were in
agreement with experimental total cross sections \cite{1,2}, specially at
low energies for He, Ne, Ar and H$_2$. Moreover, these studies yielded
correct results for resonance and binding energies for the S
wave electronic
singlet state of Ps-H \cite{18,7} and 
Ps-Li \cite{15a} systems in addition to experimental pick-off quenching
rate in
Ps-He \cite{6x} scattering.

In the present work we use the above exchange potential to study Ps-Na and
Ps-K scattering using the
three-Ps-state coupled channel method. 
We find resonances in the singlet channel  at low energies in S,
P and D waves 
 of both systems near the  Ps(2)
excitation threshold. We also report angle-integrated  elastic 
and Ps-excitation cross
sections for both systems.

The appearance of resonances  
in electron-atom \cite{cal} and positron-atom
\cite{po}
scattering, and in other atomic processes in general, is of great
interest.
Several resonances in the  
electron-hydrogen
system  have been found in  the close-coupling calculation and later
reconfirmed in the variational calculation \cite{eh}. Resonances have also
been found in the close-coupling calculation of electron scattering by
Li, Na
and K \cite{ealk}.  These resonances provide the
necessary testing ground for a theoretical formulation, which can
eventually be detected experimentally. Detailed dynamical description of
the important degrees of freedom in a theoretical formulation is necessary
for the appearance of these resonances. The ability of the present  
exchange potential to reproduce the resonances in diverse Ps-atom
systems \cite{7,15a} assures its realistic nature.

The theory for the coupled-channel study of Ps  scattering
with the regularized model potential has already appeared in the
literature \cite{6,7,15} and we quote the relevant
working
equations here. For target-elastic 
scattering 
we solve the following 
Lippmann-Schwinger scattering integral
equation in momentum space \begin{eqnarray} f^\pm_{\nu ',\nu} (
{\bf k',k})&=& {\cal B}^\pm _{\nu ',\nu}({\bf k ',k})  \nonumber
\\ &-&\sum_{\nu ''} \int \frac{ {\makebox{d}{\bf k''}} } {2\pi^2}
\frac { {\cal B}^ \pm _ {\nu ', \nu''} ({\bf k ',k''}) f^ \pm
_{ \nu'' ,\nu} ({\bf k'',k}) } {{k}^2_{\nu ''}/4-k''^2/4+
\makebox{i}0} \label{4} \end{eqnarray} where the singlet (+) and triplet
($-$)
``Born"
amplitudes, ${\cal B}^\pm$, are given by $
 {\cal B}^\pm_{\nu ',\nu}({\bf k',k}) =
 g^D_{\nu ',\nu}({\bf
k',k})\pm g^ E_{\nu ',\nu}({\bf k',k}),$
   where $g^D$ and $g^E$ represent the direct and exchange Born amplitudes
and the $f^ \pm$ are the singlet and triplet scattering amplitudes,
respectively. The quantum states are labeled by the indices $\nu$,
 referring to the Ps atom. The
variables ${\bf k}$, ${\bf k'}$, ${\bf k''}$ etc denote the appropriate
momentum states of Ps; ${\bf k}_{\nu ''}$ is the on-shell relative
momentum of Ps in the channel $\nu ''$.  We use atomic
unit (a.u.) where $\hbar = m = 1$ with  $m$ is the electron mass.

To avoid
complication
of calculating exchange potential with a many-electron wave function,
we consider a frozen-core one-electron
approximation for the targets Na and K. Such wave
functions have been successfully used for scattering of 
alkali metal atoms in other contexts and also for positronium scattering
by Li \cite{5}.
The Na(3s) and K(4s) frozen-core hydrogen-atom-like wave functions are
taken  as 
\begin{eqnarray}\label{na}
\phi_{\mbox{Na}}({\bf r})= \frac{1}{9\sqrt 3\sqrt{4\pi\bar
a_0^3}}(6-6\rho+\rho^2)e^{-\rho/2}
\end{eqnarray}
\begin{eqnarray}\label{k}
\phi_{\mbox{K}}({\bf r})= \frac{1}{96\sqrt{4\pi\bar
a_0^3}}(24-36\rho+12\rho^2-\rho^3)e^{-\rho/2}
\end{eqnarray}
 where $\rho=2r\alpha $ with 
$\alpha=1/(n\bar a_0)$. Here    $n=3$ for Na and = 4 for K and $\bar 
a_0=(2n^2E_i)^{-1}a_0$ with $E_i$ the ionization energy of the target in
a.u. and $a_0$ the Bohr radius of H. Here we use
the following experimental values for  ionization energies for Na and
K, respectively: 5.138 eV and 4.341 eV \cite{hg}.

The  direct Born  amplitude of Ps scattering is given by \cite{6,6a}
\begin{eqnarray}\label{1x}
g^D_{\nu',\nu} ({\bf k_f,k_i})&= &
\frac{4}{Q^2}
\int \phi^*({\bf r})\left[ 1-\exp ( \makebox{i} {\bf Q.
r})\right]\phi({\bf r})
\makebox{d}{\bf r}\nonumber \\ &\times&
\int \chi^*_{\nu '}({\bf  t })2\mbox{i} \sin ( {\bf Q}.{\bf  t
}/
2)\chi_{\nu}({\bf  t } ) \makebox{d}{\bf  t },
\end{eqnarray}
where $\phi({\bf r})$ is the target wave function and $\chi({\bf t})$ is
the Ps wave function.
The exchange amplitude corresponding to the  model potential   
is given by \cite{7} \begin{eqnarray}\label{1}
g^E_{\nu',\nu} ({\bf k_f,k_i})&= &
\frac{4(-1)^{l+l'}}{D}
\int \phi^*({\bf r})\exp ( \makebox{i} {\bf Q. r})\phi({\bf
r})
\makebox{d}{\bf r}\nonumber \\ &\times&
\int \chi^*_{\nu '}({\bf  t })\exp ( \makebox{i}{\bf Q}.{\bf  t }/
2)\chi_{\nu}({\bf  t } ) \makebox{d}{\bf  t }
\end{eqnarray} with
\begin{equation}\label{pa}
D=(k_i^2+k_f^2)/8+C^2[\alpha^2+(\beta_\nu^2+
\beta_{\nu'}^2)/2]
\end{equation}
where $l$  and $l'$ are  the angular momenta of the initial and final Ps
states and $C$ is the only parameter of the exchange potential. The
initial and
final Ps momenta are ${\bf k_i}$  and ${\bf k_f}$, ${\bf Q = k_i -k_f}$,
  and $\beta_\nu^2$
and $\beta_{\nu '}^2$ are the binding energies of the initial and final
states of   Ps in a.u.,
respectively.  It has been demonstrated for the Ps-H system that
at high energies  the model-exchange amplitude (\ref{1})
reduces to \cite{am} the Born-Oppenheimer exchange amplitude \cite{19}.
This exchange potential
for Ps scattering is considered
\cite{6}  to be a
generalization 
of the Ochkur-Rudge exchange potential for electron scattering \cite{20}.

After a partial-wave projection, the system of coupled equations (\ref{4}) 
is solved by the method of matrix inversion. 
 Forty Gauss-Legendre quadrature points are
used in the discretization of
each momentum-space integral.  
The calculation is performed with the
exact Ps wave functions and frozen-core orbitals (\ref{na}) and (\ref{k})
for Na and K ground
state. We consider  Ps-Na and Ps-K  scattering using the
three-Ps-state model 
that includes the 
 following states: 
Ps(1s)Na(3s), Ps(2s)Na(3s), Ps(2p)Na(3s), and 
Ps(1s)K(4s), Ps(2s)K(4s),
Ps(2p)K(4s), for Na and K, respectively.  The parameter $C$ of the
potential given by (\ref{1}) and (\ref{pa}) was adjusted to fit the
accurate theoretical results \cite{mit} for PsNa and PsK binding
energies which
are 0.005892 a.u. and 0.003275 a.u., respectively. We find
that $C=0.785$ fits both binding energies well in the three-Ps-state
model and this value of $C$ is used in all calculations reported here.
The proper strength of the model potential is obtained 
by fitting the binding energies of the Ps-Na and Ps-K systems, and it is
expected that this choice of $C$ would lead to a good overall description
of scattering in these systems.   
We recall that this value of $C$ also reproduced the accurate variational
result of PsH resonance energy in a recent five-state model for Ps-H
scattering \cite{7}. A similar variation of the parameter $C$ from unity
also led to a good overall description of the total scattering cross
section in agreement with experiment \cite{2} in the Ps-He system
in the energy range 0 eV to 70 eV \cite{6a}.

The Ps-Na and Ps-K systems have an effective attractive interaction in
the electronic singlet channel as in Ps-H \cite{7} and Ps-Li \cite{15a}
systems. The targets of these systems have one active electron outside a
closed shell.  
In the present three-Ps-state calculation we  find resonances in the
singlet channel in both systems.
No resonances appear in the triplet channel possibly because of a
predominantly repulsive interaction in this channel.
For the resonances to appear, the inclusion of the excited states of Ps is
fundamental in a coupled-channel calculation.  The static-exchange model
with both the target and Ps in the ground state does not lead to these
resonances. A detailed study of these resonances in coupled-channel study
of Ps-H \cite{18,7} and Ps-Li \cite{15a} systems in the singlet channel
has appeared in the literature. 

Here to study the resonances, first we calculate the S-, P- and D-wave
elastic phase shifts and cross sections in the singlet channel of the
Ps-Na and Ps-K systems
using the 3-Ps-state model.  In figure 1 we show the low-energy singlet
S-wave cross sections. 
The singlet S-wave phase shifts near the
resonance
energies 
are shown in the off-set of figure
1. The Ps-Na system has a  resonance at 4.58 eV of width 0.002 eV. The
resonance in the Ps-K system appears at 4.77 eV and  has a width 0.003
eV.  The phase shift curves 
clearly show the resonances where the phase shifts jump by $\pi$.

In figure 2 we show the singlet P-wave Ps-Na and Ps-K elastic phase
shifts; the corresponding singlet P-wave cross sections 
are shown in the off-set. Both
systems possess resonances at 5.07 eV of width of 0.3 eV. The
cross sections 
clearly exhibit these resonances.  In figure 3 we plot the D-wave singlet
elastic cross sections for Ps-Na and Ps-K systems at low energies. There
is
a structure in both systems at 5.3 eV which is more diffuse than in S and
P waves. 

Next we calculate the different partial cross sections of Ps-Na and Ps-K
scattering. The convergence of the cross sections with respect to partial
waves is slower in
this case than in the case of Ps-H scattering. At a incident Ps energy of
50 eV, 40 partial waves were used to achieve convergence of the
partial-wave scheme. In figures 4 and 5 we plot different angle-integrated
partial cross sections of Ps-Na and Ps-K scattering, respectively.
Specifically,
we plot the elastic, Ps(2s) and Ps(2p) excitation cross sections using the
three-Ps-state model. For comparison we also plot the elastic cross
section obtained with the static-exchange model. The elastic cross section
is  large at low energies in both systems.  The
effect of the inclusion of highly polarizable Ps(2) states in the coupling
scheme could be considerable, specially at low energies. The local minima
in the three-Ps-state elastic cross section for both systems at about $4
\sim 5$ eV are 
manifestations of the P- and D-wave resonances in this energy region.
Similar minima found in electron scattering by alkali-metal atoms 
are also consequences of  resonances \cite{ealk}.

To summarize, we have performed a three-Ps-state coupled-channel
calculation of Ps-Na and Ps-K scattering at low energies using a
regularized symmetric nonlocal electron-exchange model potential 
\cite{6} successfully used \cite{6,7,6a,6x,15,16,15a} previously in
different Ps scattering problems. The only parameter of the model
potential was adjusted to fit accurate theoretical result for PsNa and PsK
binding \cite{mit}.  We present the results of
angle-integrated partial cross
sections at different Ps energies.  
 We find resonances in S, P and D waves
near the Ps(2) excitation threshold.
In this study we have used a three-Ps-state model. Similar resonances have
been observed in the coupled-channel study of electron-H \cite{eh},
electron-Na, electron-K \cite{ealk}, positron-hydrogen \cite{po}, Ps-H
\cite{18,7} and Ps-Li \cite{15a} systems. In most cases, a more complete
calculation and (in some cases) experiments have reconfirmed these
resonances. Hence we do not believe that the appearance of
resonances in the present three-state calculation to be so peculiar as to
have no general validity. On the contrary, in view of the correlation
found
between resonance and binding energies in the singlet Ps-H system
\cite{7}, it is expected that the  reproduction of correct binding
energies of the Ps-Na
and Ps-K systems in the present model should lead to correct resonance
energies in these systems.
 However, the resonance energies might change
slightly after a more complete calculation (with accurate many-body wave
functions of the target and including the excited states of the target)
and it would be
intersting to study the present resonances using more complete theoretical
models in the future in addition to compare the present results with
future experiments.

The work is supported in part by the Conselho Nacional de Desenvolvimento -
Cient\'\i fico e Tecnol\'ogico,  Funda\c c\~ao de Amparo
\`a Pesquisa do Estado de S\~ao Paulo,  and Finan\-ciadora de Estu\-dos e
Projetos of Brazil.

\newpage
{\bf Figure Caption:}

1. Singlet S-wave elastic cross sections  at different Ps energies for
Ps-Na
(dashed
line) and Ps-K (full line) scattering. 
The corresponding phase shifts
near resonance are shown in the off-set.

2. Singlet P-wave elastic phase shifts   at different Ps energies for
Ps-Na
(dashed
line) and Ps-K (full line) scattering. The corresponding cross sections
are shown in the off-set.

3. Singlet D-wave elastic cross sections  at different Ps energies for
Ps-Na
(dashed line) and Ps-K (full line) scattering. 

4. Partial cross sections for Ps-Na scattering at different Ps energies:
three-Ps-state elastic (full line), three-Ps-state Ps(2s) (dashed-dotted
line), three-Ps-state Ps(2p) (short-dashed line), static-exchange elastic
(long-dashed line).

5. Same as in figure 4 for Ps-K scattering.

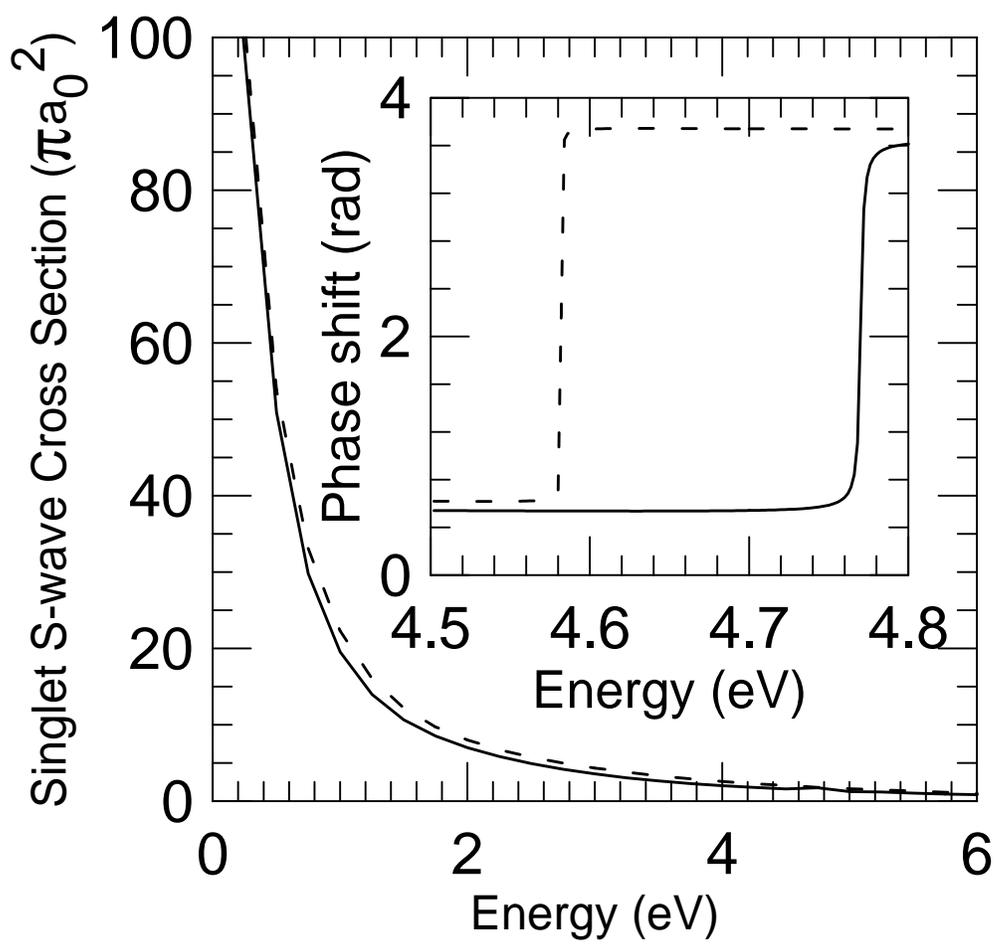
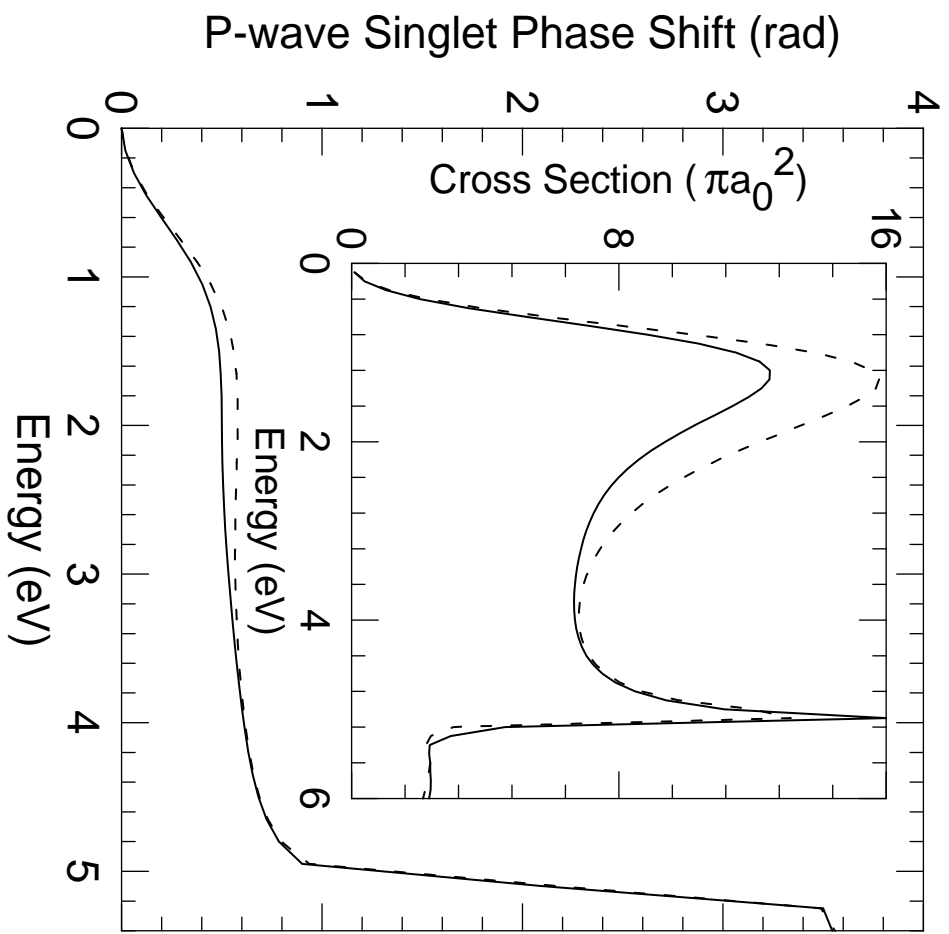
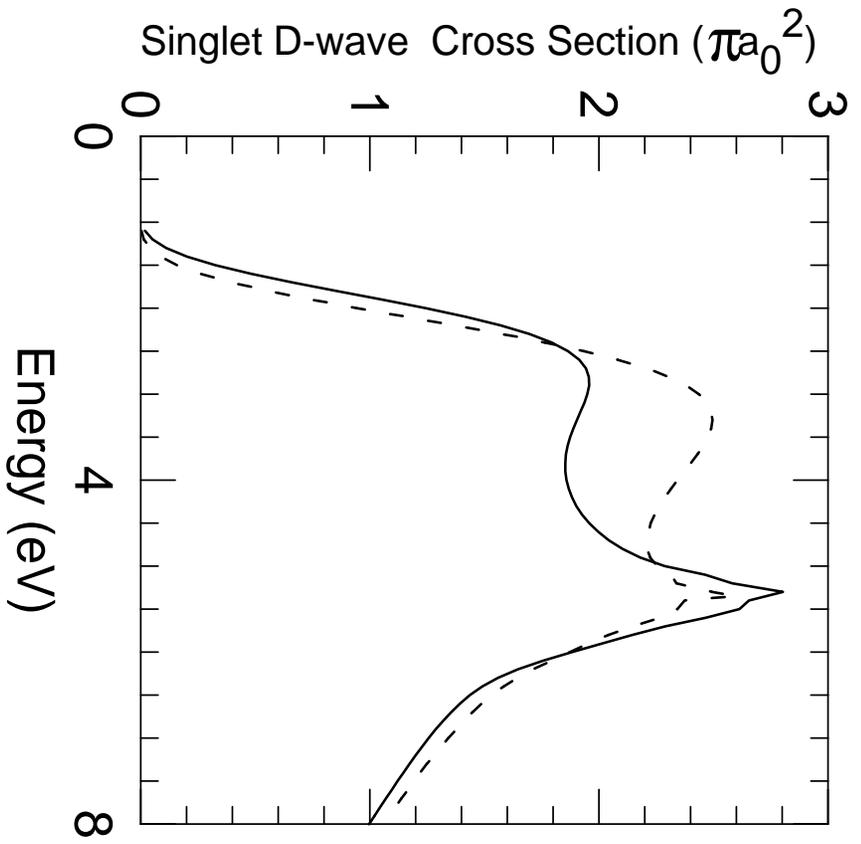
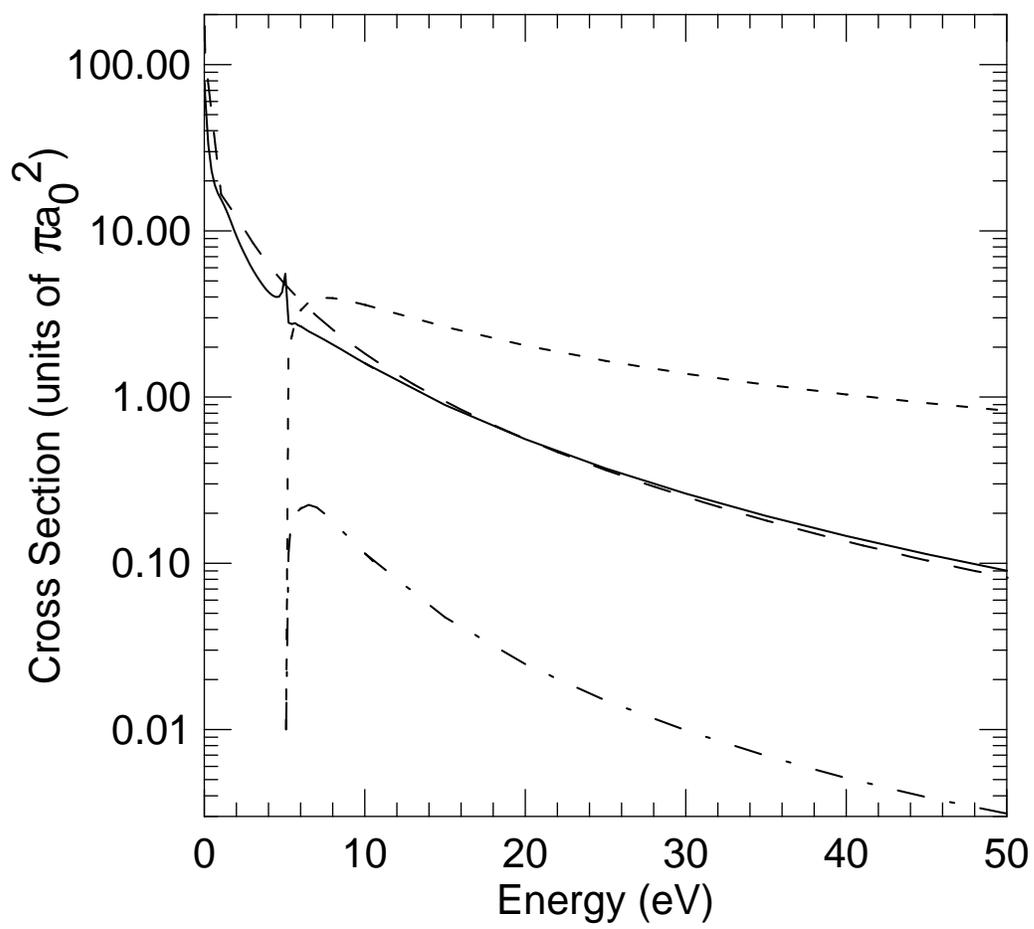
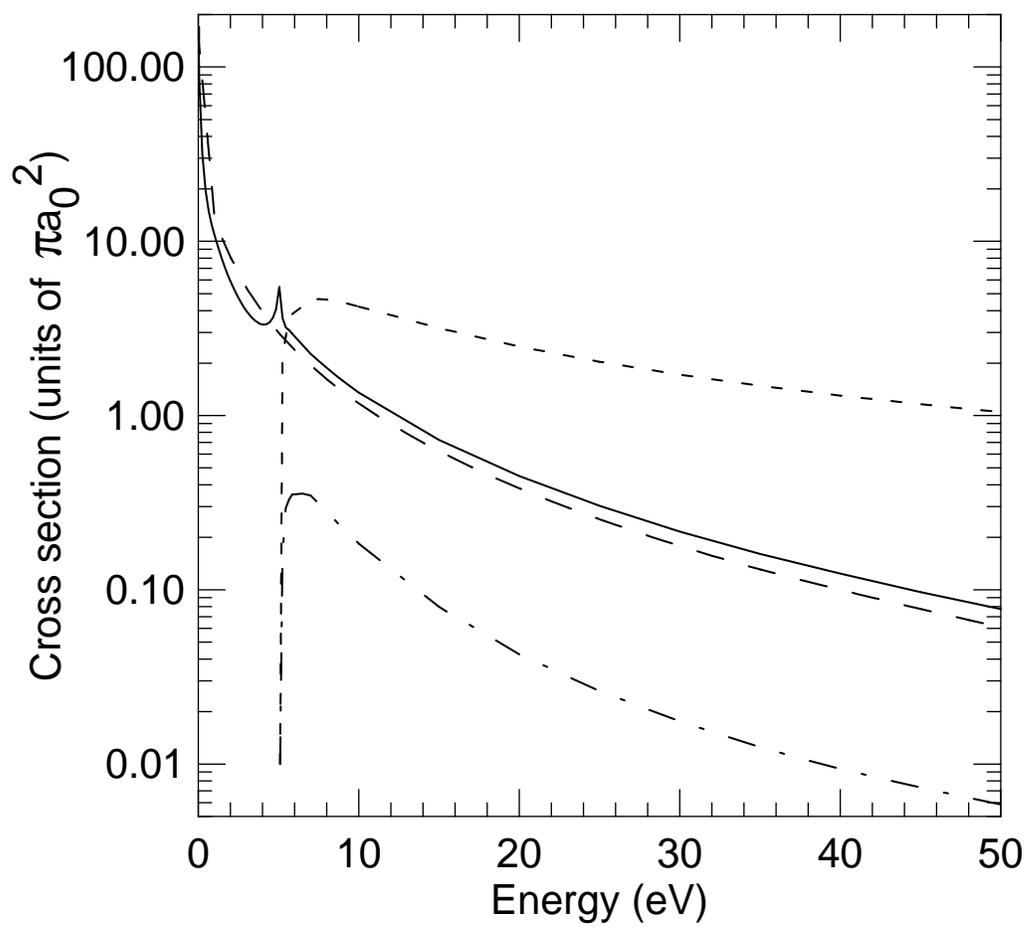
\end{document}